\documentclass[12pt,preprint]{aastex}

\shorttitle{$\emph{Interaction of Two Active Region Filaments}$}
\shortauthors{Yang et al.}

\begin{document}

\title{Interaction of Two Active Region Filaments Observed by NVST and SDO}

\author{Liheng Yang\altaffilmark{1,2,3}, Xiaoli Yan\altaffilmark{1,3}, Ting Li\altaffilmark{2,3}, Zhike Xue\altaffilmark{1,3}, and Yongyuan Xiang\altaffilmark{1,3} }

\altaffiltext{1}{Yunnan Observatories, Chinese Academy of Sciences, Kunming 650216, China; yangliheng@ynao.ac.cn}

\altaffiltext{2}{Key Laboratory of Solar Activity, National Astronomical Observatories, Chinese Academy of Sciences, Beijing 100012, China}

\altaffiltext{3}{Center for Astronomical Mega-Science, Chinese Academy of Sciences, Beijing, 100012, China}

\begin{abstract}

Using high spatial and temporal resolution H$\alpha$ data from the New Vacuum Solar Telescope (NVST) and
simultaneous observations from the Solar Dynamics Observatory (SDO), we present a rare event on the interaction
between two filaments (F1 and F2) in AR 11967 on 2014 January 31. The adjacent two filaments were
almost perpendicular to each other. Their interaction was driven by the movement of F1 and started when
the two filaments collided with each other. During the interaction, the threads of F1
continuously slipped from the northeast to the southwest, accompanied by the brightenings at the junction of two filaments and
the northeast footpoint of F2. Part of F1 and the main body of F2 became invisible in H$\alpha$ wavelength due to the heating
and the motion of F2. At the same time, bright material initiated from the junction of two filaments were observed to move
along F1. The magnetic connectivities of F1 were found to be changed after their interaction. These observations
suggest that magnetic reconnection was involved in the interaction of two filaments and resulted in the eruption of one filament.

\end{abstract}


\section{Introduction}

Filaments are relatively cool and dense plasma suspended in the hot corona. They are usually
observed to form and be maintained in a magnetic channel \citep{mar98}. Recent observations show that
the material of filaments are supported by magnetic flux ropes, which can be visible when filaments are activated or
erupted \citep{che11,li13a,li13b,li13c,yan14,yan15}. Filaments are known to have magnetic patterns
of handness or chirality \citep{mar98}. Viewing the filament from the positive-polarity side,
the chirality is defined to be dextral if the filament axial field is directed rightward and sinistral if directed
leftward. Generally, filaments follow a handedness rule, in which dextral filaments dominate in the
north hemisphere and sinistral filaments dominate in the south one \citep{mar94}. It is believed that the filament chirality
and the magnetic helicity sign have a one-to-one correspondence, in which dextral filaments have negative
(left-handed) magnetic helicities and sinistral filaments have positive (right-handed) helicities \citep{cha00}.

Previous observations reveal that two nearby filaments can approach and interact with each
other \citep[e.g.,][]{aza48,kon13,li12,xue16a}. Some authors have presented observations that filaments of the same chirality,
sharing a common filament channel, could merge into a single filament at their neighboring
ends \citep{den00,sch04,bon09,jos14,zhu15}. While \citet{su07} showed that two sinistral filaments along
different channels could merge by body coalescence and form a common filament channel.
More recently, \citet{jia14} reported that the intrusion of the erupted material of one sinistral
filament into the filament channel of the other sinistral filament caused the merge of the two
filaments. Unlike the above observations, \citet{kum10} and \citet{cha11} found that two
dextral filaments observed on 2003 November 20 first approached each other, then merged at their middle
parts, and finally separated in opposite directions. Consequently, the two filaments changed their
footpoint connections to form two newly linked filaments. Similar observations were also reported
by \citet{jia13}.

Under convective zone conditions, \citet{lin01} simulated the collision of two highly twisted, identical flux ropes. In their numerical
simulation, four types of interaction were found on the basis of the helical sign and contact angle of
the two flux ropes. They were bounce, merge, slingshot, and tunnel, in which the magnetic reconnection
was involved. The interaction was called bounce when the two collided flux tube bounced off each other,
and was called merge when they merged into one. The slingshot interaction occurred when the two flux ropes with opposite
helicities collided with each other in an appropriate contact angle. As a result, two new flux ropes with exchanging
footpoint connections formed and then slingshotted away from the interaction region. In the following work of \citet{lin06},
the collision of two lowly twisted and unequal flux ropes with the same helicity can also produce the slingshot interaction.
Under coronal conditions, \citet{tor11} also found the slingshot interaction of two flux ropes with the same helicity,
which was based on the event that occurred on the 2003 November 20. Note that the opposite and same helicities mentioned above
refer to the sign of helicity. The tunnel and slingshot interaction had the same initial condition.
Nevertheless, the slingshot reconnected once at the collision site and finally formed two new flux tubes. The tunnel
reconnected twice at two different places and exchanged the section between the two reconnection points, making the flux tubes to
pass through each other. Among them, the slingshot interaction was the most energetic because the magnetic flux
was annihilates and twist canceled. In addition, \citet{dev05} performed numerical simulations of the formation and
interaction between pairs of head-to-tail filaments within the sheared-arcade model. Four possible basic combinations
of chiralities (identical or opposite) and axial magnetic fields (aligned or opposed) between the filaments have been considered.
The numerical experiments suggest that tether-cutting reconnection occurred between filaments with aligned axial fields,
irrespective of their chiralities.

There exists mainly two mechanisms that lead to the instability of filaments and trigger the energy release and eruptions.
One is the non-ideal process such as magnetic reconnection \citep{lin00,ant99,moo01,dev08,she12,li16,xue16b}, and the other is an ideal
magnetohydrodynamic (MHD) process such as kink and torus instabilities \citep{tor05,kli06,sri10,yana,bi15}. Observations
have indicated that magnetic reconnection during the filament interaction can lead to filament eruptions. \citet{kim01} observed a
filament that partially erupted after a rapid change of connectivity in a bundle of filament threads. \citet{su07} pointed out that one
filament eruption resulted from the sudden mass injections produced by external bodily magnetic reconnection between two filaments.
More recently, \citet{che16} reported that tether-cutting reconnections between filaments triggered filament eruptions.

To our knowledge, observational evidences on filament interaction are relatively rare and need to be further studied,
which is helpful to understand physical mechanisms of such interaction. In this paper, we present
detailed analysis of the interaction of two filaments with the high-resolution, multi-wavelength observations
from the New Vacuum Solar Telescope \citep[NVST;][]{liu14} and the Solar Dynamics Observatory \citep[SDO;][]{pes12}.
We describe the observational data in Section 2 and present analysis results in Section 3. Conclusions and discussions were
given in Section 4.

\section{Observations and Data Analysis}

The NVST has a multi-channel high resolution imaging system, which is used to observe
the fine structures in the photosphere and the chromosphere. Up to now, it can image the Sun in
three channels, i.e., TiO, G-band, and H$\alpha$. The H$\alpha$ channel is centered
at 6562.8~{\AA}, with a bandwidth of 0.25~{\AA}. It can provide off-band observation
in the range of $\pm$5~{\AA} with a step size of 0.1~{\AA}. In the present work,
the H$\alpha$ line center images from 06:55:00 UT to 07:59:59 UT on 2014 January 31
were employed to analyze the interaction of the two filaments in the chromosphere.
These images have a pixel size of 0.162$''$ and a cadence of 12~s.
The field-of-view (FOV) is 150$''$$\times$150$''$. The H$\alpha$ data are first processed by dark current subtracted
and flat field corrected, and then are reconstructed by speckle masking \citep{wei77,loh83}.

The Atmospheric Imaging Assembly \citep[AIA;][]{lem11} on board SDO provides full-disk
images taken in seven EUV passbands and three continuum bands. The spatial resolution and cadence of these images
are 1.5 $''$ and 12~s, respectively. Here, all of the seven EUV channels were used, which cover
a temperature range from 0.05 MK to 20 MK. The differential emission measure (DEM) using the
almost simultaneous observations of six AIA EUV lines (131~{\AA}, 94~{\AA}, 335~{\AA}, 211~{\AA},
193~{\AA}, and 171~{\AA} formed at coronal temperature) was reconstructed, and the DEM-weighted average
temperature was calculated \citep{che12}. We also used the Helioseismic and Magnetic Imager (HMI) \citep{sch12}
line-of-sight magnetograms and vector magnetic field data from the SDO. The line-of-sight magnetograms were used to
extrapolate the potential fields above the photosphere, which were used to calculate the decay index of magnetic
fields overlying on the F2. They have a cadence of 45~s and a sampling of 0.5 $''$ pixel$^{-1}$. The vector
magnetic field data were obtained by using the Very Fast Inversion of the Stokes Vector algorithm \citep{bor11}. The 180 azimuthal
ambiguity was resolved based on the minimum energy method \citep{met94,met06,lek09}. These data have a pixel size of 0.5 $''$ pixel$^{-1}$ and
a cadence of 12 minutes. We processed the AIA images and HMI line-of-sight magnetograms by using the standard routine
aia$\_$prep.pro in the SolarSoftWare (SSW) packages \citep{fre98}, and rotated all the images to a reference
time (07:30 UT on 2014 January 31). In addition, the Global Oscillation Network Group (GONG) full-disk H$\alpha$ image
was used to align the SDO and NVST data.

\section{Results}

On 2014 January 31, the observations of the NVST covered the interaction process of two filaments (``F1'' and ``F2''), which were located close
together in a complex NOAA AR 11967 (S14E41). Figure 1 shows their locations outlined by the blue dashed lines in the SDO/AIA 304~{\AA} and H$\alpha$ images
with the SDO/HMI radial component of the vector magnetic field superimposed before F1 began to move. The black and white contours indicate the negative and positive magnetic field. The level of the contours is $\pm$100 G. It is clear that F1 is roughly oriented along the east-west direction, while F2 oriented along the north-south direction. The axes of the two filaments were nearly perpendicular to each other (see Figures 1(a) and (b)). When viewed at the positive-polarity side, the axial field direction of F2 was leftward. Therefore, F2 is a sinistral filament, which is consistent with the preferential filament pattern in the southern hemisphere \citep{mar94}. It can be seen from Figure 1(b), F1 was not strictly located on the polarity inversion line (PIL), but its location was close to the PIL. It is supposed that F1 might be suspended high in the corona, and the misalignment
between F1 and the PIL was due to the projection effect.

Figure 2 displays the interaction of the two filaments and the subsequent filament eruptions in H$\alpha$, 304~{\AA} and
335~{\AA} wavelengths, respectively. Prior to the interaction, the two filaments can be identified as dark structures not only
in the cool channels such as H$\alpha$ and AIA 304~{\AA} channels, but also in the hot channels such as AIA 335~{\AA} channel
(see Figures 2(a) -- (a2)). At about 06:55 UT, the central part of F1 was activated and began to move to the southwest direction
(see Animation 1). As a result, F1 gradually approached F2 from the northeast to the southwest direction (see Figures 2(b) -- (b2)). Such approaching motion can be clearly
seen in the stack plots along the slit ``A -- B'' , in which the two filaments can be identified as dark tracks (see Figures 5(b) -- (d)).
At about 07:22 UT, F1 impinged on the northeast part of F2 and the brightenings appeared at the junction of the two filaments. The brightenings
were enhanced with further filament interaction, which were marked by the white arrows in Figures 2(c) -- (c2). Simultaneously with the
appearance of the brightenings, the cool plasma of filament was heated up and started to move along F1. The moving heated plasma was observed as
bright structures in Figures 2(c1), (c2), (d1), and (d2), which were marked by the thin white arrows. It was exhibited as many bright stripes in the
stack plots along the slit ``C -- D'' (see Figure 5(e)). At the same time, part of F1 and the main body of F2 became invisible in the H$\alpha$ wavelength (see Figure 2(c)), while they exhibited as obvious brightenings at 304~{\AA} wavelength (see Figure 2(c1)). The disappearance of F1 might be mainly due to the heating of the filament plasma, but that of F2 might have two probabilities. one is the heating of the filament plasma, and the other is the motion of F2. Unfortunately, NVST has no off-band observations on that day. At about 07:25 UT, F2 started to erupt. Meanwhile, it underwent an anticlockwise rotation during its eruption. Three minutes later, F2 collided with its overlying large-scale loops and began to interact with them, as shown in Figure 3. These large-scale loops were brightened by the interaction and can be clearly seen in Figure 3(b). As the interaction was processed, the overlying large-scale loops were continually stretched and the material of F2 was found to be falling down along them (see Figure 3(c) and Animation 2). The interaction led to the deflection of F2. F2 finally developed to a fan structure (as marked by the thick white arrows in Figures 2(d) -- (d2)). At about 07:28 UT, two flare ribbons appeared on opposite sides of the eruptive F2 (marked as ``FR'' in Figures 2(d) -- (d2)). It was too weak to be recorded as a flare, similar to observations shown by \citet{yan08}. Several minutes later, another pair of flare ribbons appeared at the east of F1 in the same active region. It was related to a GOES C2.3 flare, which started at 07:35 UT and peaked at 07:57 UT. No coronal mass ejection was associated with this eruption.

In order to present the filament interaction in more detail, we extracted the interaction region (as marked by the white square in Figure 2(b))
and analyzed its evolution process at H$\alpha$ and 304~{\AA} wavelengths, as showed in Figures 4(a) -- (l). The southwest border of F1 was
traced out by red dashed curves in Figures 4(a) -- (g). Owing to the high spatial resolution of the NVST, the filament threads can be clearly
distinguished (see Figures 4(a) -- (h)). It is noticed that the F1's threads extended to the east before its interaction with the threads of F2, as shown in Figure 4(a).
At about 07:22 UT, the southwest bunch of F1's threads collided with the northeast bunch of F2's threads. It seems that the two bunches of threads
joined together (see Animation 3). One minute later, they began to separate and F1's threads continued to move to the southwest direction (see red dashed lines in Figures 4(b) -- (h)).
Note that the bunch of F1's threads that was involved in the filament interaction changed their directions, and curved to the southwest direction, indicating that the
magnetic connectivities of F1's threads have changed. As the interaction continued, the magnetic connectivities of F1's threads changed continuously, exhibiting as
a slipping motion from the northeast to the southwest direction  (see Animation 3). The F1's threads finally rooted in the weak network region, where the
northwest footpoint of F2 was originally located (see Figure 4(h)). The newly formed magnetic structure was depicted by the yellow dashed line in Figure 4(h). We also noticed that the brightenings appeared at the northwest footpoint of F2 when the two filaments collided with each other (see Animation 3). Following the movement of F1's threads, the brightenings were gradually strengthened and extended to the northwest direction, which also showed up as a slipping motion (see thick red arrows in Figures 4(b) -- (g) ). Such kind of brightenings was ever observed by \citet{kim01} in a pre-eruption reconnection event, and was considered to be caused by the downward draining filament material produced by the magnetic reconnection. And more notably, the brightenings appeared at 304~{\AA} wavelength when the F1's threads collided with the F2's threads, as shown by the yellow arrow in Figure 4(j). They were enhanced as the interaction in progress, and can be clearly identified in chromospheric and coronal lines (see Figures 4(k), (m) -- (o) and Animation 3). The temperature map obtained with the DEM method was shown in Figure 4(p), and it is found that the temperature of the brightenings can reach up to 20 MK. These observations suggest that magnetic reconnection might occur between F1 and F2 during the interaction process. The brightenings at the northwest footpoint of F2 indicate that magnetic reconnection during the filament interaction altered the stable environment of F2. We speculate that the downward magnetic tension force might decreased when magnetic reconnection occurred, which might trigger the eruption of F2.

To investigate the brightenings at the junction of the two interactive filaments, we selected a small region marked by the white box
in Figure 4(j) to calculate the variation of H$\alpha$, SDO/AIA 304~{\AA} and 335~{\AA} brightness over time, as shown in Figure 5(a).
We note that these curves started to increase at about 07:22 UT (marked by the yellow vertical line), which was the time when the two filaments
encountered each other. For tracing the eruption of F1 before interaction, we made time slices from H$\alpha$, 304~{\AA}, and 335~{\AA} images
along a slit ``A -- B'' (marked by a white line in Figure 2(a)). As seen from the Figures 5(b) -- (d), the filaments appeared as dark tracks. Clearly,
F1 was slowly approaching F2 at a speed of 3 -- 5 km\,s$^{-1}$. During this process, F2 almost did not move. In order to trace the mass motions
along F1, we made a time slice from AIA 304~{\AA} images along a slit ``C -- D'' (marked by a white line in Figure 2(b1)). It is noted that these
mass motions started at about 07:22 UT, and had a velocity of 150 km\,s$^{-1}$.

In order to better understand the eruption of F2 in this event, the magnetic topology and the decay index over the filament
were obtained by performing the local potential field extrapolation \citep{ali81,gar89}. The decay index is defined
as n=-$\emph{d}$ln($\emph{B}$)/$\emph{d}$ln($\emph{h}$), where $\emph{B}$ is the transverse strength of the magnetic field and $\emph{h}$ is
the height measured from the photosphere. Here decay indexes were derived from a central cross section of F2 (marked by the red line in Figure 6(a))
at a height range from 0.9~Mm to 22~Mm. The results were shown in Figure 6(b). Here, the height of F2 was estimated. We assumed that F2 has a loop
shape vertical to the solar surface and its footpoints rooted in the same atmospheric level. The line of sight has a 37$^\circ$ angle relative to the solar surface.
So the maximum height of F2 is estimated to be about 12.9 Mm, which was marked by the blue dashed line in Figure 6(b). The red line in Figure 6(b) denotes
where the cross section cut of F2. As seen from Figure 6(a), one can see that F2 was restricted by the overlying large-scale arcades (marked by the golden field
lines in Figure 6(a)). From Figure 6(b), we note that F2 was located on the intersection point of the red solid line and the blue dashed line, and the decay
index above F2 was lower than 0.3, which was far below the theoretical threshold value (1.5) of torus instability \citep{kli06}. Therefore, torus instability is not
the mechanism of F2's eruption. However, F2 erupted during the filament interaction, which suggests that F2's eruption was mainly caused by the filament interaction.

\section{Conclusions and Discussion}

The interaction of two adjacent filaments (F1 and F2) was well observed by the NVST and SDO at the southeast of the Sun
on 2014 January 31. F1 and F2 were close and almost perpendicular to each other. The interaction was driven by the movement of F1,
which had a velocity of 3 -- 5 km\,s$^{-1}$. The interaction occurred when F1 and F2
collided with each other and produced brightenings at the junction of these two filaments. The threads of F1 continuously slipped from
the northeast to the southwest, accompanied by the brightenings at the northwest footpoint of F2. Moreover, the motions of bright material can be
seen along F1 from the junction of the two filaments. The temperature of the brightenings at the junction of the two filaments was up to 20 MK.
During the interaction, part of F1 and the main body of F2 became invisible in H$\alpha$ wavelength but appeared as evident brightenings in the 304~{\AA} wavelength,
which might be due to the heating and the motion of F2. The interaction led to the change of the magnetic
connectivities of F1's threads. Finally, F2 erupted after their interaction.

The interaction was driven by the F1's movement. Previous observations show that there are two kind of mechanisms driving the filament interaction.
One is slow photospheric motions \citep{kum10,cha11,tor11} and the other is the eruption of one of the interactive filaments \citep{jia13,jia14,kon13}.
It is obvious that the filament interaction in this event was driven by the second mechanism.

Magnetic reconnection normally occurred during the filament interaction. Based on the observations of the converging motions of the two filaments,
the subsequent appearance of a hot plasma layer, and a coronal hard X-ray source near the interaction interface, \citet{zhu15} pointed out that the two filaments
reconnected with each other during their interaction. In this event, evidences of magnetic reconnection in the filament interaction region were also presented,
including high-temperature brightenings (up to 20 MK) at the junction of the interactive filaments,
mass motions along F1, significant plasma heating, the magnetic connectivity change of F1, and brightenings at the northwest footpoint of F2.
During the filament interaction, the connectivities of F1's threads changed continuously, accompanied by the slipping motion of the brightenings at
the northeast footpoint of F2. To our knowledge, this is the first time to display the details of filament interaction process, especially the magnetic
connectivity change of filament threads, which was owing to the high temporal and spatial resolution of the NVST.

In this event, no clear observational signatures for bounce, merge or tunnel were found. However, it is noted that F1's magnetic connectivities
changed after the interaction, indicating that at least one new filament formed. Therefore, the interaction of this event might be inclined to support
the slingshot interaction. This event was associated with a GOES C2.3 flare. Since the eastern leg of F1 was rooted in between the two flare ribbons and the
interaction changed the magnetic connectivities of F1 and F2, this flare might be resulted from the eruption of F2.

During the filament interaction, F2 became unstable and erupted. Previous observations show that magnetic reconnection during filament
interaction can initiate solar eruptions. \citet{su07} present observations that the overloaded mass ejection caused by the magnetic reconnection
during the filament interaction led to the second filament eruption. \citet{bon09} pointed out that the build-up shear of long linking flux tubes
by magnetic reconnection involved in the filament interaction , together with the removal of overlying field by magnetic flux reconnection, led to
the instability and eventual eruption of the interactive filaments. Very recently, \citet{che16} claimed that the tether-cutting reconnection between
two filaments triggered solar eruptions. In this case, it is also magnetic reconnection involved in the filament interaction that might trigger F2's eruption.
According to the result of torus instability, F2 should be stable before the filament interaction. It was balanced under the upward magnetic pressure force,
downward magnetic tension force, and downward gravity force. The downward magnetic tension force might be decreased when magnetic
reconnection occurred between the two filaments, which breaks the balance of F2's strapping field, and leads to the final eruption of F2.

\acknowledgments

The authors thank the referee for her/his constructive suggestions and comments, which help to improve this manuscript.
The authors would like to thank Prof. Jingxiu Wang, Prof. Jun Zhang, Prof. Yunchun Jiang, Dr. Yi Bi, and Dr. Jincheng Wang for their
useful discussions. The authors are indebted to the NVST and SDO science teams for providing the wonderful data. This work is supported
by the National Natural Science Foundations of China (11373066, 11303050, 11303049 and 11203037),
the Youth Project of Western Light of Chinese Academy of Sciences, the Talent Project of Western Light
of Chinese Academy of Sciences, Yunnan Science Foundation of China under number 2013FB086, Youth Innovation Promotion Associated CAS (no. 2011056),
and the Key Laboratory of Solar Activity of CAS (KLSA201407 and KLSA201412).

\appendix

\clearpage

\begin{figure}
\epsscale{1.0}
\plotone{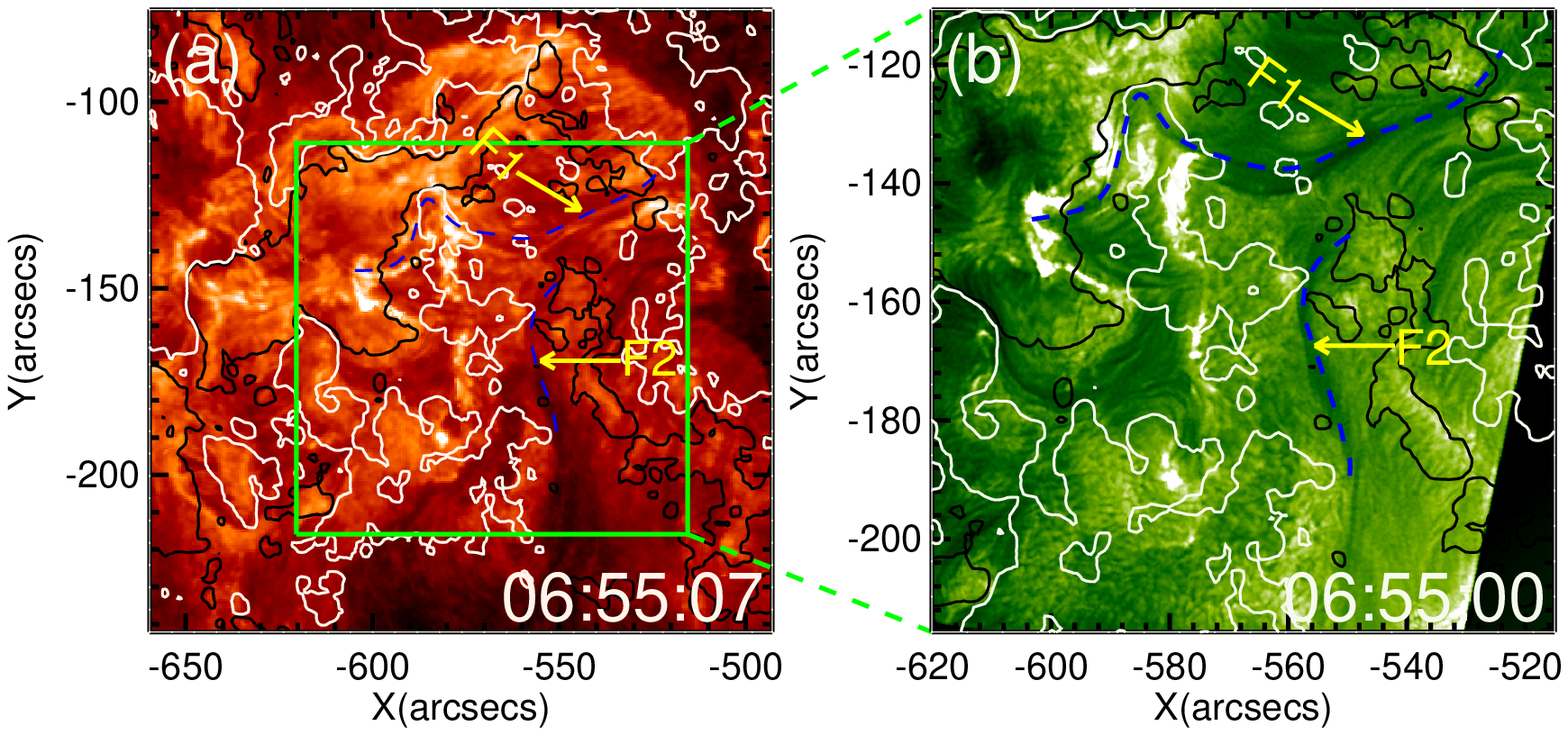}
\caption{SDO/AIA 304~{\AA} (a) and NVST H$\alpha$ (b) images overplotted by the contours of the SDO/HMI
radial component of the vector magnetic field showing the locations of the active region filaments. ``F1'' and ``F2'' represent the
two interactive filaments. The two filaments are delineated by the blue dashed lines. Black/white contours represent
positive/negative polarity regions, and the contour level is $\pm$100 gauss. The field of view (FOV) of panel (b) is
outlined by the green square in panel (a). \label{fig1}}
\end{figure}

\clearpage

\begin{figure}
\epsscale{0.8}
\plotone{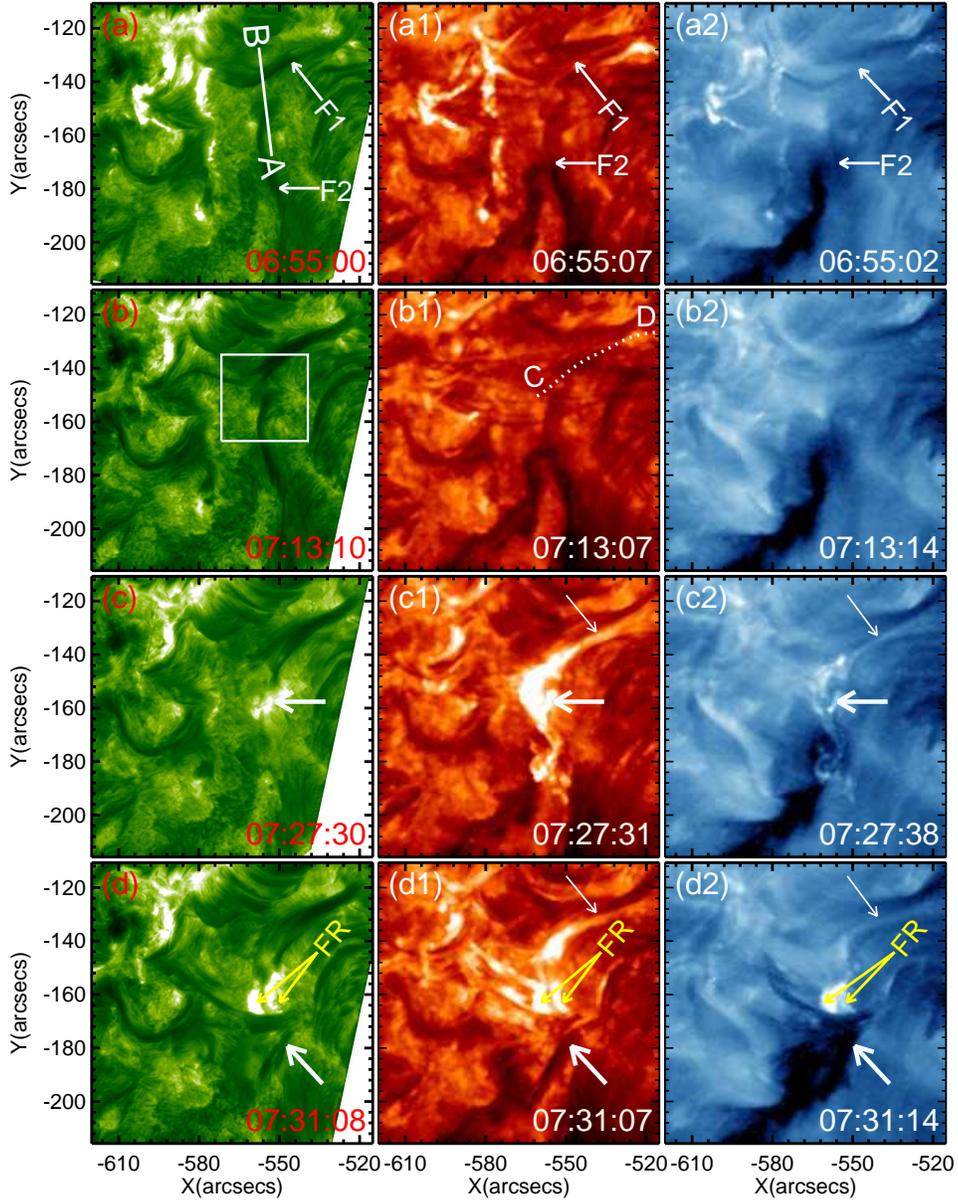}
\caption{NVST H$\alpha$ (a -- d), SDO/AIA 304~{\AA} (a1 -- d1), and 335~{\AA} (a2 -- d2) images showing the interaction of the two
filaments and subsequent filament eruptions. The line ``A'' -- ``B'' marks the slit position of space-time plots of
panels (b) -- (d) in Figure 5. The curve line ``C'' -- ``D'' marks the slit position of the space-time plot of
panel (e) in Figure 5. ``FR'' represents the flare ribbons involved in this event. The white square in panel (b) gives
the FOV of Figure 4. The thick white arrows in panels (c) -- (c2) point to the erupting F2. The thin white arrows
in panels (c1), (c2), (d1), and (d2) point to the mass motions. The thick white arrows in panels (d) -- (d2)
indicate the fan structure of F2 after its eruption. \label{fig2}}
\end{figure}

\clearpage

\begin{figure}
\epsscale{1.0}
\plotone{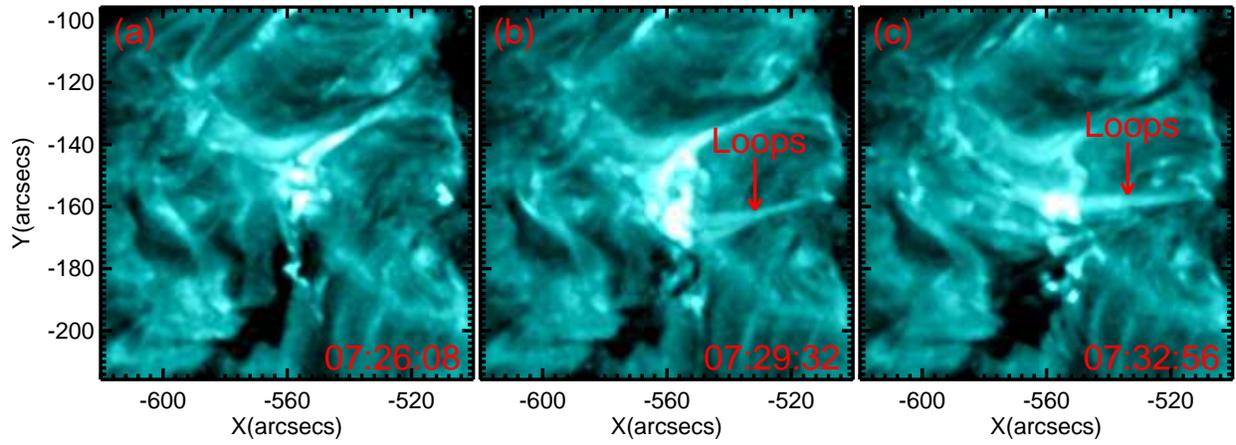}
\caption{SDO/AIA 131~{\AA} (a -- c) images showing the interaction of the eruptive filament and its overlying loops.
The red arrows in panels (b) and (c) indicate the overlying loops. \label{fig3}}
\end{figure}

\clearpage

\begin{figure}
\epsscale{1.0}
\plotone{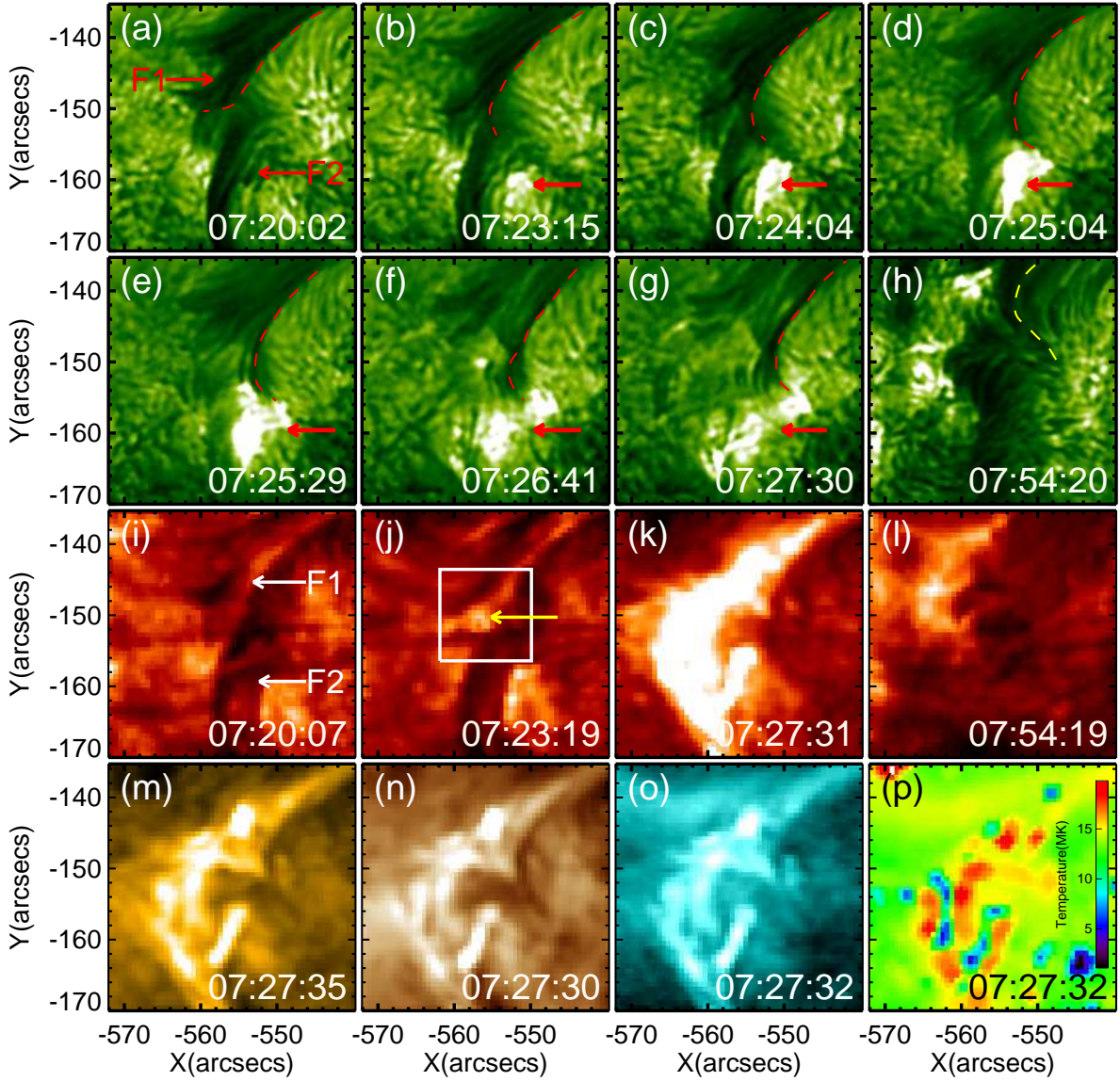}
\caption{Sequences of NVST H$\alpha$ (a -- h) and SDO/AIA 304~{\AA} (i -- l) images
showing the interaction of the two filaments. SDO/AIA 171~{\AA} (m), 193~{\AA} (n), 131~{\AA} (o), and temperature (p) images
displaying the interaction region. The red dashed lines depict the southwest border of F1. The yellow dashed line represents the newly formed magnetic structure.
The thick red arrows indicate the brightenings at the northwest footpoint of F2. The white square in panel (j) is used to calculate the changes of brightness during the filament interaction. The yellow arrow in panel (j) represents the brightenings during the filament interaction. \label{fig4}}
\end{figure}

\clearpage

\begin{figure}
\epsscale{0.9}
\plotone{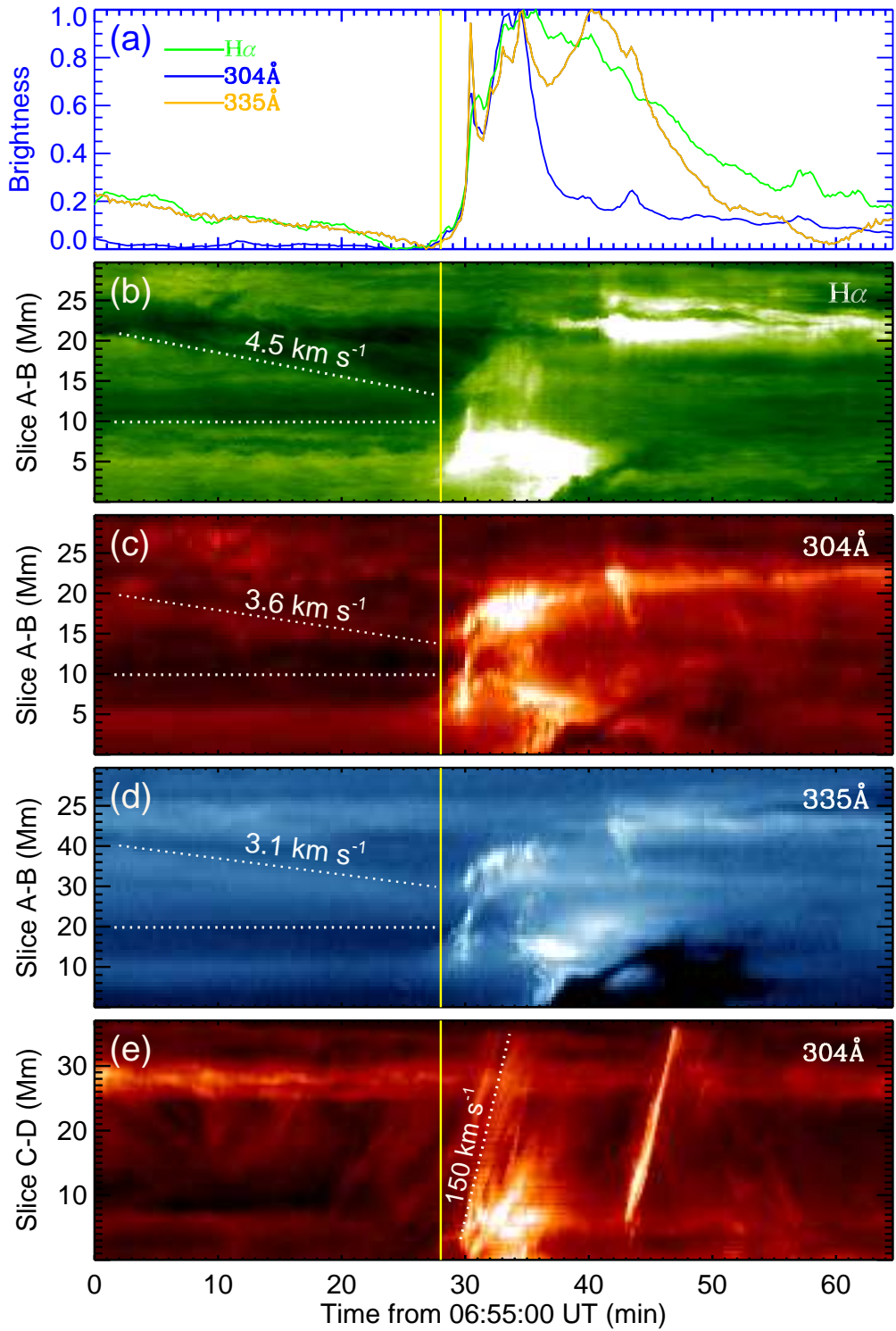}
\caption{(a) Time profiles of NVST H$\alpha$, SDO/AIA 304~{\AA}, and 335~{\AA} brightness in
a region marked by the white square in Figure 4(j). The light curves are normalized to one.
(b) -- (d) Space-time plots along the line ``A'' -- ``B'' marked in Figure 2(a). (e) the space-time plot along
the curve ``C'' -- ``D'' marked in Figure 2(b1). \label{fig5}}
\end{figure}

\clearpage

\begin{figure}
\epsscale{1.0}
\plotone{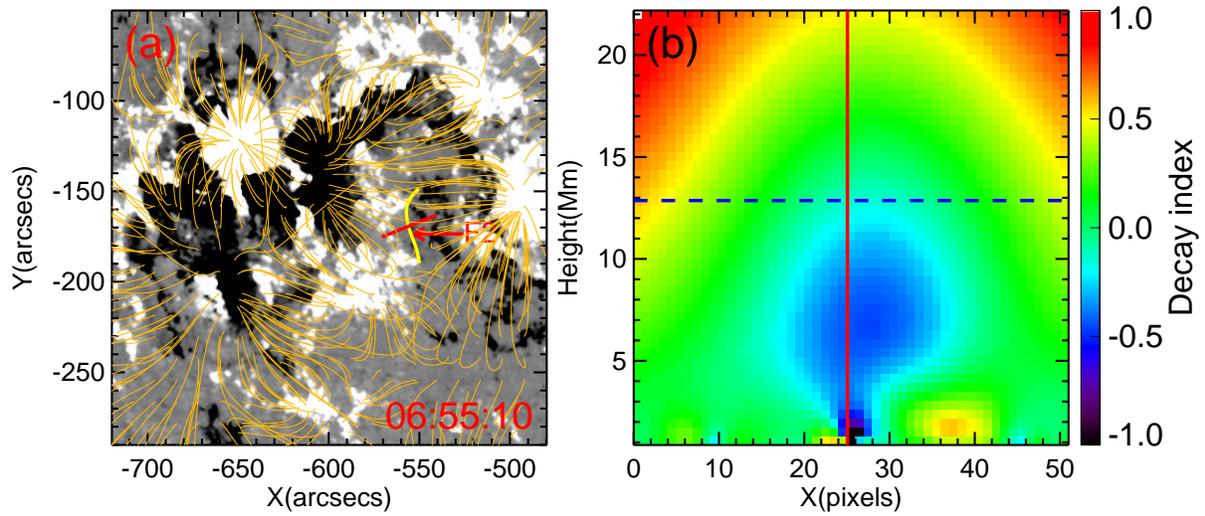}
\caption{(a) SDO/HMI line-of-sight magnetogram with the extrapolated field lines superimposed,
along with the pre-interaction F2. (b) Decay indexes measured in the cross section across the F2
marked by the red line in panel (a). The red line in panel (b) marks where the cross section cut of F2.
The blue dashed line in panel (b) denotes the measured height of F2. \label{fig6}}
\end{figure}

\end{document}